\documentclass{article}
\usepackage{fullpage,amsthm,amsfonts,amsmath,times}

\newtheorem{theorem}{Theorem}
\newtheorem{lemma}{Lemma}
\newtheorem{corollary}{Corollary}
\newtheorem{proposition}{Proposition}

\newcommand{\ket}[1]{\lvert #1 \rangle} 
 
\newcommand{\norm}[1]{\lVert #1 \rVert} 
 
\newcommand{\size}[1]{\lvert #1 \rvert}

\newcommand{\eps}{\varepsilon}
\newcommand{\floor}[1]{\lfloor #1 \rfloor}
\newcommand{\set}[1]{\left\{ #1 \right\}}

\newcommand{\uc}{\textsc{Unique Collision}}
\newcommand{\gc}{\textsc{Group Commutativity}}
\newcommand{\usc}{\textsc{Unique Split Collision}}

\newcommand{\suppress}[1]{}
\newcommand{\journal}[1]{#1}

\begin{document}

\title{Quantum Complexity of\\ Testing Group Commutativity\thanks{%
A preliminary version of this paper appeared in
\emph{Proceedings of 32nd International Colloquium on Automata, Languages and Programming},
Lecture Notes in Computer Science, pages 1312-1324, Springer-Verlag,
Berlin, 2005.\medskip
}}
\author{Fr\'ed\'eric Magniez\thanks{
CNRS-LRI, France.
Email : \texttt{magniez@lri.fr}.
Address: LRI - b\^{a}timent 490,
Universit\'e Paris-Sud,
91405 Orsay cedex,
France.
Partially supported by
the EU 5th framework program RESQ IST-2001-37559, and by ACI
Cryptologie CR/02 02 0040 and ACI S\'ecurit\'e Informatique 03 511
grants of the French Research Ministry.  Part of the research was done
while visiting Perimeter Institute at Waterloo, ON, Canada.}
\and
Ashwin Nayak\thanks{
University of Waterloo and Perimeter Institute for Theoretical Physics,
Canada.
Email:  \texttt{anayak@math.uwaterloo.ca}.
Address: Department of Combinatorics and Optimization
and Institute for Quantum Computing, University of Waterloo,
200 University Ave.\ W., Waterloo, Ontario N2L 3G1, Canada.
Research supported in part by NSERC, CIAR, MITACS, CFI, OIT (Canada) and
ARDA (USA).
Research at Perimeter Institute for Theoretical Physics is supported
in part by the Government of Canada through NSERC and by the Province
of Ontario through MRI.
}
}

\date{May 7, 2007}

\maketitle

\begin{abstract}
We consider the problem of testing the commutativity of a black-box
group specified by its $k$ generators. The complexity (in terms of~$k$) of this problem was first considered by Pak, who
gave a randomized algorithm involving~$O(k)$
group operations. We construct a quite optimal quantum algorithm for this
problem whose complexity is in~$\tilde{O}(k^{2/3})$.  
The algorithm uses and highlights the power of the quantization method of
Szegedy.
For the lower bound of~$\Omega(k^{2/3})$, we 
give a reduction from a special case of Element Distinctness to our problem.
Along the way, we prove 
the optimality of the algorithm of Pak for the
randomized model.
\end{abstract}

\section{Introduction}

A direction of research in quantum computation pioneered by
Grover~\cite{gro96} around search problems in unstructured,
structured, or partially structured databases has recently been
infused with new ideas for algorithm design.  In contrast to problems
based on the Hidden Subgroup Problem (HSP) (see for instance
Ref.~\cite{mos99}), the speed up for these search problems is often
only polynomial.

Usually in search problems, the access to the input is done via an
oracle.  This leads to the notion of query complexity which measures
the number of accesses to the oracle.  While no significant lower
bounds are known for quantum time complexity, the oracle constraint
sometimes enables us to prove such bounds in the query model. For
promise problems quantum query complexity indeed can be exponentially
smaller than the randomized one. A prominent example is HSP.  On the
other hand, for total functions, deterministic and quantum query
complexities are polynomially related~\cite{bbcmw01}.

In 
HSP, 
the group with its all structure is
known to the algorithm designer, and the group operations are
generally efficiently computable. In the event that the group is not
explicitly known, or the group operations are not efficient to
implement, it is appropriate to model the group operations by an
oracle or a black-box.  The notion of {\em black-box groups} was
introduced by Babai and Szemer\'edi~\cite{bs84}.  In this
model, the elements of a group 
are encoded by words over a finite
alphabet, and the group operations are performed by an oracle (the
black-box). The groups are assumed to be specified by generators, and
the encoding of group elements is not necessarily unique: different
strings may encode the same group element.  
Mosca~\cite{mos99} showed that one can learn in quantum polynomial
time the structure of any black-box abelian group. Such a task is
known to be hard classically.  
Then Watrous~\cite{wat01} pioneered the study of
black-box group properties in the quantum context. 

In this context, we study the problem of testing the commutativity of a
black-box group (\textsc{Group Commutativity}) given by its
generators. The classical complexity of this problem was first
considered by Pak~\cite{pak00}.
The straightforward
algorithm for the problem has complexity~$O(k^2)$, where~$k$ is the
number of generators, since it suffices to check if every pair of
generators commute. Pak presented a surprising randomized algorithm
whose complexity is linear in~$k$, and also showed that the
deterministic lower bound is quadratic. The linear upper bound on
complexity may also be obtained by applying quantum
search~\cite{gro96} to locate a pair of generators that do not
commute. Using the quantization of random walks by
Szegedy~\cite{sze04}, we instead present a {\em sublinear\/} algorithm
with time and query complexity
in $\tilde{O}(k^{2/3})$ (\textbf{Theorem~\ref{algo-thm}}), where the
$\tilde{O}$ notation means that logarithmic multiplicative factors are
omitted.

\textsc{Group Commutativity} bears a deceptive resemblance to
\textsc{Element Distinctness}.  The aim in the former is to detect the
presence of a pair of generators which collide in the sense that they
do not commute.  However, since the group structure is unknown,
whether or not a pair of generators collide can only be determined by
invoking the group oracle.  Moreover, the group oracle provides access
to elements from the entire group spanned by the given generators,
which may be used towards establishing commutativity.  These
differences necessitate the use of ideas from Pak's algorithm, the
theory of rapidly mixing Markov chains, and perhaps most remarkably,
the Szegedy quantization of walks.

\textsc{Group Commutativity}
appears to be the first natural problem for which the approach of
Szegedy has no equivalent using other known techniques for
constructing quantum algorithms, such as Grover search~\cite{gro96},
or the type of quantum walk introduced by Ambainis~\cite{amb04}.
Conversely, for Triangle Finding, the approach of Ambainis
was more successfully applied. For this problem, Magniez, Szegedy and
Santha~\cite{mss05} construct a quantum algorithm that uses
recursively two quantum walks {\em {\`a} la\/} Ref.~\cite{amb04}, while
the Szegedy quantization of walks seems to give a less query-efficient
algorithm.  The problems of \textsc{Group Commutativity} and
\textsc{Triangle Finding} thus give strong evidence that the walks due
to Ambainis are not comparable with the ones due to Szegedy.

A recent result of Buhrman and \v{S}palek~\cite{BuhrmanS06} on matrix product
verification also relies on the Szegedy quantization for its
worst case {\em time\/} complexity. However, for the worst case instances,
when there is at most one erroneous entry, the approach of Ambainis
gives an algorithm whose query complexity is the same as that due to
Szegedy.

We also prove that our algorithm is almost optimal by giving an
$\Omega(k^{2/3})$ lower bound for the quantum query complexity of
\textsc{Group Commutativity}
(\textbf{Theorem~\ref{slb-thm}}). Simultaneously, we give an
$\Omega(k)$ lower bound for its randomized query complexity
(\textbf{Theorem~\ref{slb-thm}}).  This lower bound shows that the
algorithm of Pak~\cite{pak00} is optimal, and to our knowledge is new.
We prove the lower bounds using a reduction from the problem of
detecting a unique collision pair of a function, which is a special
case of \textsc{Element Distinctness}.

\section{Preliminaries}

\subsection{Black-box groups}

We suppose that the elements of the group $G$ are encoded by
binary strings of length $n$ for some fixed integer $n$,
which we call
the {\em encoding length\/}.  The groups are given by generators,
and therefore the {\em input size} of a group is the product of the
encoding length and the number of generators.  For simplicity, we also
assume that the identity element of the group is given.  Note that the
encoding of group elements need not be unique,
i.e., a single group element
may be represented by several strings. If the encoding is not unique,
one also needs an oracle for identity tests. Unless otherwise
specified, we assume that the encoding is unique in this paper.  All
of our results apply when the encoding is not unique if one is given
an oracle for identity tests.

Since we deal with black-box groups we shall shortly describe
them in the framework of quantum computing (see also
Refs.~\cite{mos99} or
\cite{wat01}).  For a general introduction to quantum computing the
reader might consult
Refs.~\cite{nc00,ksv02}.  We work in the quantum
circuit model.  For a group $G$ of encoding length $n$, the black-box
is given by two oracles $O_G$ and its inverse $O_G^{-1}$, both
operating on $2n$ qubits.  For any group elements $g,h \in G$, the
effect of the oracles is the following:
$\quad O_G\ket{g}\ket{h} = \ket{g}\ket{gh} \quad\text{and}\quad
O_G^{-1}\ket{g}\ket{h} = \ket{g}\ket{g^{-1}h}$.
In this notation we
implicitly use the encoding of a group element. We do that
everywhere in the paper when there is no ambiguity. 
Not every binary string of length $n$ necessarily corresponds to a group
element.  In this case the behaviour of the black-box can be arbitrary.

\subsection{Query model}

The quantum query model was 
explicitly introduced by Beals, Buhrman, Cleve,
Mosca, and de Wolf~\cite{bbcmw01}.  In this model, as in its classical
counterpart, we pay for accessing the oracle, but unlike the classical
case, the machine can use the power of quantum parallelism to make
queries in superposition.

The state of the computation is represented by three registers, the
query register~$g$, the answer register~$h$, and the work register
$z$. The computation takes place in the vector space spanned by all
basis states $\ket{g,h,z}$.  In the {\em quantum model} the state of
the computation is a complex combination of all basis states which has
unit length in the~$\ell_2$ norm.

\suppress{ In the randomized model it is a non-negative real
combination of unit length in the norm $l_1$, and in the deterministic
model it is always one of the basis states.  }

For a black-box group the query operator is $O_G$ together with
its inverse $O_G^{-1}$. For oracle function~$F:X\to Y$ the query
operator is $O_F : \ket{g}\ket{h}\mapsto \ket{g}\ket{h\oplus F(g)}$,
where $\oplus$ denotes the bitwise xor operation. 

Non-query operations are independent of the oracle.  A {\em $k$-query
algorithm\/} is a sequence of $(k+1)$ operations $(U_0, U_1, \ldots ,
U_k)$ where each $U_i$ is unitary.
\suppress{ in the quantum and stochastic in the randomized model, and
it is a permutation in the deterministic case.  }
Initially the state of the computation is set to some fixed value
$\ket{\bar{0},\bar{0},\bar{0}}$. 
In case of an oracle function,
the sequence of operations $U_0, O_F, U_1, O_F, \ldots, U_{k-1}, O_F,
U_k$ is applied.  For black-box groups, the modified sequence of
operations $U_0, O_G^{b_1}, U_1, O_G^{b_2}, \ldots, U_{k-1},
O_G^{b_k}, U_k$ is applied,
where~$b_i \in \{\pm 1\}$.  Finally, one or more
qubits designated as output bits are measured to get the outcome of
the computation.  
The quantum algorithms we
consider
have a probabilistic outcome, and they
might give an erroneous answer with non-zero probability.
However, the probability of making
an error is bounded by some fixed constant~$\gamma < 1/2$.

In the query model of computation each query adds one to the {\em
query complexity} of an algorithm, but all other computations are
free.
The {\em time complexity\/} of the algorithm is usually measured in
terms of the total circuit size for the unitary operations~$U_i$.  We
however take a more coarse-grained view of time complexity, and
assume that operations such as accessing qubits containing group
encodings or updating them, take unit time.

\subsection{Quantum walks}
\label{sec-qwalks}

We state a simple version of the recent result of
Szegedy~\cite{sze04}.  Let~$P$ be an irreducible (i.e., strongly
connected), aperiodic (i.e., non-bipartite), and symmetric Markov
chain on a graph $G=(V,E)$ on~$N$ vertices. Such a walk is necessarily
ergodic, i.e., converges to a unique stationary distribution
regardless of the initial state.

Let~$P[u,v]$ denote the transition probability from $u$ to $v$.  Let
$M$ be a set of marked nodes of $V$.  Assume, one is given a database
$D$ that associates some data $D(v)$ to every node $v\in V$.  {From}
$D(v)$ we would like to determine if $v\in M$. We expedite this search
using a quantum procedure $\Phi$.  When operating with $D$ three types
of cost are incurred. The cost might denote any measure of complexity
such as query or time complexities.

\suppress{ all measured in the number of queries to the oracle.  Using
arguments of~\cite{hmw03}, we can also deal with an error-bounded
procedure $\Phi$.  Nevertheless for the sake of clarity, we assume in
the rest of this section that $\Phi$ is a zero-error procedure.
\begin{description}\setlength{\itemsep}{0pt}
}
\noindent
\textbf{Setup cost $\mathsf{S}$:} The cost to set up $D(v)$ for a $v\in V$.\\
\textbf{Update cost $\mathsf{U}$:} The cost to update $D(v)$ for a
$v\in V$, i.e., moving from $D(v)$ to $D(v')$, where the transition
from $v$ to $v'$ is allowed by the Markov chain $P$.\\
\textbf{Checking cost $\mathsf{C}$:} For~$v \in V$, the complexity of
checking if~$v \in M$ from~$D(v)$.

Concerning the quantization of the walk $P$, one needs to consider the
quantum time complexity of its implementation in terms of the
following parameters:

\noindent
\textbf{Initialization time $\mathsf{I}$}: The time complexity for
constructing the 
superposition
$$\frac{1}{\sqrt{N}} \sum_{u,v} \sqrt{P[u,v]} \ket{u,v}.$$
\textbf{Transition time $\mathsf{T}$:} The time complexity of
realizing the 
transformation
\[
\ket{u,v} ~~\mapsto~~  2\, \sqrt{P[u,v]} \sum_{v'}\sqrt{P[u,v']}
\ket{u,v'} - \ket{u,v}.
\]
The Markov chains we construct in this paper are all random
walks on regular graphs. For every node~$u$, the
probabilities~$P[u,v]$ are all equal to~$1/d$ or~$0$, where~$d$ is the
degree of each node in the graph. The unitary transformation defined
above, restricted to the node~$u$ in the first register, then
corresponds to the Grover diffusion operator~\cite{gro96} on the
neighbours of~$u$. The diffusion operator
is the unitary matrix~$\frac{2}{d}\mathbb{J} - \mathbb{I}$.

In the following theorem, which is the main result of
Ref.~\cite{sze04}, the notation $O(\cdot)$ denotes the existence of a
universal constant so that the expression is an upper bound.
\begin{theorem}[Szegedy \cite{sze04}]\label{mario-thm}
Let $\delta$ be the eigenvalue gap of $P$, and let
$\frac{\size{M}}{\size{V}} \ge \eps > 0$ whenever~$M$ is non-empty.
There exists a quantum algorithm that determines if $M$ is non-empty
with cost
$\mathsf{S}+O({(\mathsf{U}+\mathsf{C})}/{\sqrt{\delta\eps}})$, and an
additional time complexity of
$\mathsf{I}+O(\mathsf{T}/\sqrt{\delta\eps})$.
\end{theorem}
Note that in this theorem, when the cost denotes the time complexity,
we need to add the additional time complexity term to it.

Szegedy's theorem thus gives us a recipe for constructing and
characterizing the behaviour of a quantum walk algorithm by specifying a
classical random walk, and analysing its spectral gap and stationary
distribution.

\subsection{Spectral gap of Markov chains}

The spectral gap (or eigenvalue gap) of a Markov chain (with
non-negative eigenvalues) is the
difference between the largest and the second largest
eigenvalue of the probability transition matrix that represents it.
Estimating this quantity directly from a description of the matrix is
often very difficult. We take an indirect route to estimating this
quantity by appealing to its relation with the convergence properties of
the Markov chain.

Consider an ergodic Markov chain on state space~$X$ with stationary
distribution~$\pi$.  Let~$P^t_x$ be the probability distribution
on~$X$ obtained by performing~$t$ steps of the Markov chain starting
at~$x$.  Let~$\Delta(t)$ be the maximum over all starting states~$x
\in X$ of the total variation distance~$\norm{P^t_x - \pi}$.  Then
the {\em mixing time\/}~$\tau$ of the Markov chain is defined as the
smallest~$t$ such that~$\Delta(t') \le \frac{1}{2\mathrm{e}}$ for
all~$t' \ge t$.

A {\em coupling\/} 
for a Markov chain is a stochastic process on pairs of
states~$(U_t,V_t)$ such that~$U_t$ and~$V_t$, viewed marginally, each
evolve according to the Markov chain, and if~$U_t = V_t$,
then~$U_{t+1} = V_{t+1}$. The 
{\em coupling time\/}~$T$ 
is the maximum
expected time (over all pairs of initial states~$(u,v)$) for the
states~$U_t, V_t$ to coincide:
$$
T = \max_{u,v}
\mathrm{E}[ \mathrm{argmin}_t \{ U_t = V_t, U_0 = u, V_0 = v \}].
$$

We use the following facts about the mixing of Markov chains:
\begin{enumerate}
\item \cite[Proposition~2.2, Chapter~2]{sinclair93} For walks with
only non-negative eigenvalues, $\lambda^t \le \Delta(t) \cdot (\min_u
\pi(u))^{-1}$, where~$\lambda$ is the second largest eigenvalue. This
bounds the second largest eigenvalue in terms of the total variation
distance.
\item (see e.g., 
Ref.~\cite{Aldous82}) 
$\Delta(t) \le 2\,
\exp(-\floor{\frac{t}{\tau}})$. This relates the total variation
distance at any time~$t$ to the mixing time~$\tau$.
\item \cite{Griffeath78} $\tau \le 2\mathrm{e} T$. This bounds the
mixing time~$\tau$ in terms of the coupling time~$T$.
\end{enumerate}

Combining all three relations, we may deduce the following relationship
between the spectral gap of a Markov chain and coupling time.
\begin{corollary}
\label{thm-gap-coupling-time}
For any ergodic Markov chain with only non-negative eigenvalues, the
spectral gap~$1-\lambda ~~\geq~~ \frac{1}{4\mathrm{e}T}$, where~$\lambda$
is the second largest eigenvalue, and~$T$ is the coupling time for 
any valid coupling defined on~$X \times X$.
\end{corollary}
\begin{proof}
Chaining all three facts listed above, taking $t$-th roots, and letting~$t
\rightarrow \infty$, we see that
\begin{eqnarray*}
\lambda & \le &  \exp(-\frac{1}{2\mathrm{e}T}) \\
        & \le & 1 - \frac{1}{4\mathrm{e}T},
\end{eqnarray*}
which is equivalent to the claim.
\end{proof}

\subsection{The problems}

\suppress{ Given group elements $g_1,\ldots,g_k$ we denote the
subgroup spanned by them by $\gspan{g_1,\ldots,g_k}$.  }

Here we define the problems we are dealing with.  
The focus of the paper is on

\begin{quote}
\textsc{Group Commutativity}\\
\emph{Oracle:} Group operations $O_G$ and $O_G^{-1}$ for an encoding in $\{0,1\}^n$\\
\emph{Input:} The value of $n$ and the encoding of generators 
$g_1,\ldots,g_k$ of $G$\\
\emph{Output:} \texttt{Yes} if $G$ is
commutative, and \texttt{No} otherwise (if there are two indices $i,j$
such that $g_i g_j\neq g_jg_i$)
\end{quote}
The next problem is a special instance of a well-studied problem,
\textsc{Element Distinctness}.
\begin{quote}
\textsc{Unique Collision}\\
\emph{Oracle:} A function $F$ from $\{1,\ldots,k\}$ to $\{1,\ldots,k\}$\\
\emph{Input:} The value of $k$\\
\emph{Output:} \texttt{Yes} if there exists a unique collision
pair $x\neq y \in\{1,\ldots,k\}$ such that $F(x)=F(y)$, and \texttt{No}
if the function is a permutation
\end{quote}
This is a promise problem (or a relation) since we do not require a
definite output for certain valid oracle functions.
We also use a further specialization of the problem when $k$ is even, 
\textsc{Unique Split Collision}, 
where,
in the \texttt{Yes} instances,
one element of the colliding pair has to come from 
$\{1,\ldots,k/2\}$ and the other from $\{k/2+1,\ldots,k\}$.
We call this a {\em split\/} collision.
Note that in the positive instances of this problem, the restriction
of the function to the two intervals~$\set{1,\ldots,k/2}$
and~$\set{k/2+1,\ldots,k}$ is injective.

A beautiful application of the polynomial method gives us the optimal
query complexity of \uc.
\begin{theorem}[\cite{as04,Kutin05,Ambainis05}]\label{ed-thm}
The quantum query complexity
of \textsc{Unique Collision}
is~$\Omega(k^{2/3})$.
\end{theorem}
The original results of the works cited above refer to the more general
problem \textsc{Element Distinctness}, which requires the detection of
one or more colliding pairs.  This was proven by a randomized
reduction from the problem \textsc{Collision} which distinguishes between
a bijection and a two-to-one function.
However, the reduction is still valid for the special case we consider.
The reason is that the randomized reduction from \textsc{Collision} 
results in instances
of \uc\ with constant probability.

\suppress{\footnote{The
restriction of a two-to-one function on~$\set{1,\ldots,k}$ to a random
subset of size~$\sqrt{k}$ has exactly one collision pair with a
constant probability independent of~$k$. See the proof of
Proposition~\ref{usc-prop} for details of how such a reduction would
work; that proof contains a reduction with a similar flavour.}}

\section{A quantum algorithm for \textsc{Group Commutativity}}

We are given a black-box group $G$ with generators $g_1,\ldots,g_k$.
The problem is to decide if $G$ is abelian.  For technical reasons
(see the proof of Lemma~\ref{thm-prob}), and without loss of
generality, we assume that $g_1$ is the identity element.

We denote by $S_l$ the set of all $l$-tuples of distinct elements of
$\{1,\ldots,k\}$.  For any $u=(u_1,\ldots,u_l)\in S_l$, we denote by
$g_u$ the group element $g_{u_1}\ldots g_{u_l}$.  Not all group
elements are generated by such products of~$l$ generators. However,
the subset of group elements we get this way has properties analogous
to the entire group (see Lemma~\ref{com-lemma} below).

Our algorithm is based on the quantization of a random walk on~$S_l
\times S_l = S_l^2$.  We adapt an approach due to Pak, for which
we generalize Lemma~1.3 of Ref.~\cite{pak00} to random elements from
$S_l$. Then we show how to walk on $S_l^2$ for finding a
non-commutative element in $G$, if there is any.  We conclude
using Theorem~\ref{mario-thm}.

In this section, we let $p=\frac{l(l-1)+(k-l)(k-l-1)}{k(k-1)}$.
Observe that when $k=2l$, then $p=\frac{l-1}{2l-1}\leq\frac{1}{2}$.
Moreover, when $l=o(k)$, then $1-p=\Theta(l/k)$.

\begin{lemma}
\label{thm-prob}
Let $K\neq G$ be a subgroup of $G$. Then
$\Pr_{u\in S_l}[g_u\not\in K]  \geq \frac{1-p}{2}.$
\end{lemma}
\begin{proof}
First we fix a total order (equivalently, a permutation)~$\sigma$
of~$\{1,\ldots,k\}$, and we denote by $S_l^\sigma$ that subset of
$l$-tuples in~$S_l$ which respect the total order~$\sigma$. In other
words, $u = (u_1,\ldots,u_l) \in S_l^\sigma$ iff $\sigma^{-1}(u_i) <
\sigma^{-1}(u_{i+1})$ for all~$1 \le i < l$.  

All the sets~$S_l^\sigma$ have the same size~$\binom{n}{l}$.  Any
tuple~$u$ of distinct elements respects exactly~$n!/l!$
permutations~$\sigma$, and therefore occurs in exactly the same number
of sets~$S_l^\sigma$.  Thus picking a uniformly random element
from~$S_l$ is the same as first picking a uniformly random
permutation~$\sigma$, and then picking a random element~$u \in
S_l^\sigma$. Consequently, it is enough to prove the theorem for any
fixed order~$\sigma$. The reader may find it helpful to take~$\sigma$
to be the identity permutation to understand the idea behind the
proof.

Let~$i$ be the smallest index for which~$g_{\sigma(i)} \not\in
K$. Such an~$i$ exists since~$K\neq G$.  Recall that~$g_1$ is the
identity element.

Fix an ordered $l$-tuple~$u$ such that~$\sigma(i) \not\in u$
and~$1 \in u$.  We denote by~$v$ the ordered $l$-tuple
where~$1$ has been deleted from~$u$, and~$\sigma(i)$ has been inserted
into it at the appropriate position (that respects the total
order). Formally, if~$u = (u_1, \ldots, u_m, u_{m+1}, \ldots, u_l)$
such that~$\sigma^{-1}(u_m) < i < \sigma^{-1}(u_{m+1})$, then~$v$ is
obtained by deleting~$1$ from the~$(l+1)$-tuple~$(u_1, \ldots, u_m,
\sigma(i), u_{m+1}, \ldots, u_l)$. This mapping defines a bijection (a
perfect matching) between tuples~$u$ such that~$\sigma(i) \not\in u$
and~$1 \in u$, and tuples~$v$ such that~$\sigma(i) \in v$ and~$1
\not\in v$. Below we show that for every matched pair of tuples~$u,v$,
at least one of the group elements~$g_u, g_v$ is not in~$K$.

Consider a matched pair~$u,v$ as above.  Let~$a = g_{u_1} g_{u_2}
\cdots g_{u_m}$, and~$b = g_{u_{m+1}} \cdots g_{u_l}$. Then~$g_u = ab$
and~$g_v = a g_{\sigma(i)} b$. Note that because of the choice of~$i$,
the group element~$a \in K$. If~$g_u = ab \in K$, then~$b = a^{-1} g_u
\in K$ as well.  This means that~$g_v = a g_{\sigma(i)} b \not\in K$,
since otherwise, we would have~$g_{\sigma(i)} = a^{-1} g_v b^{-1} \in
K$. Thus, both~$g_u$ and~$g_v$ cannot be in~$K$. Therefore
\[
\Pr_{u\in S_l^\sigma} [g_u \in K | \sigma(i) \in u \text{ xor } 1 \in u]
   ~~\leq~~ \frac{1}{2}.
\]
Since for any two indices~$i,j$,
\[
\Pr_{u\in S_l^\sigma} [i,j \in u \text{ or } i,j \not\in u] 
    ~~=~~  p ~~=~~  \frac{l(l-1)+(k-l)(k-l-1)}{k(k-1)},
\]
we conclude that
\[
\Pr_{u\in S_l^\sigma} [g_u \in K] 
    ~~\leq~~ (1-p)\times\tfrac{1}{2}+p\times 1,
\]
which is at most~$(1+p)/2$.
\end{proof}
From Lemma~\ref{thm-prob}, we can generalize Lemma~1.1 of
Ref.~\cite{pak00}. For this, we recall the notions of the
centre of a group, and the centralizer of a group element.  The
centralizer~$C(g)$ of a group element~$g \in G$ is the set of all
group elements~$h$ that commute with~$g$. This is a subgroup
of~$G$. The centre~$C(G)$ of the group is the intersection of the
centralizers of all groups elements. This is also a subgroup of~$G$,
and by definition, the elements of this subgroup commute with every
element of the group~$G$.

\begin{lemma}\label{com-lemma}
If $G$ is non-commutative then
$\Pr_{u,v\in S_l} [g_u g_v \neq g_v g_u]  \geq \frac{(1-p)^2}{4}.$
\end{lemma}
\begin{proof}
If~$G$ is non-commutative, then the centre~$C(G)$ of~$G$ is a proper
subgroup.  With probability at least~$(1-p)/2$, $g_u$ does not belong
to~$C(G)$ for a random~$u \in S_l$ (Lemma~\ref{thm-prob}).  We
condition upon this event. Since~$g_u \not\in C(G)$, there is at least
one element of~$G$ that does not commute with it. So the centralizer
of~$g_u$ is also a proper subgroup of~$G$.  Again, by
Lemma~\ref{thm-prob}, the probability that for a random~$v \in S_l$,
$g_v$ does not belong to the centralizer of~$g_u$ is also at
least~$(1-p)/2$.
\end{proof}

For~$u \in S_l$,
let $t_u$ be the balanced binary tree with~$l$ leaves, whose leaves
are from left to right the elements $g_{u_i}$, for $i=1,\ldots, l$,
and such that each internal node is the group product of its two
successors. If $l$ is not a power of $2$, we put the deepest leaves to
the left.

The random walk on~$S_l^2$ that forms the basis of our quantum
algorithm
consists of two independent simultaneous walks
on~$S_l$. For a pair~$(u,v)$ of~$l$-tuples, we maintain the
binary trees~$t_u,t_v$ as described above as the data.

\begin{quote}
{\bf The random walk on~$S_l$}\\
Suppose the current state is~$u \in S_l$.\\
With probability~$1/2$ stay at~$u$; with probability~$1/2$, do the following:\\
-- Pick a uniformly random position~$i \in
\{1,\ldots,l\}$, and a uniformly random index~$j \in \{1, \ldots,k\}$.\\
-- If~$j = u_m$ for some~$m$, then exchange~$u_i$ and~$u_m$, 
else, set~$u_i = j$. \\
-- Update the tree $t_u$ (using $O(\log l)$ group operations).
This involves ``uncomputing'' all the products from the root to the leaf
that is being updated, and then computing fresh products from the leaf
to the root of the tree.
\end{quote}

\begin{lemma}
\label{thm-gap}
The spectral gap of the walk described above is at least~$\frac{c}{l
\log l}$, for a universal constant~$c \ge \frac{1}{8\mathrm{e}}$,
provided~$l \le k/2$.
\end{lemma}
\begin{proof}
First, we show that the random walk mixes rapidly using a
``coupling argument''. Then, using a relation between mixing time and
the second largest eigenvalue, we get a bound on the spectral
gap.

Note that the walk is ergodic and has the uniform distribution
on~$S_l$ as its stationary distribution~$\pi$. Thus~$\pi(u) =
\frac{(k-l)!}{k!}$ for all~$u$. 

The eigenvalues of any stochastic matrix~$P$, such as the transition
matrix of a Markov chain, all lie in the interval~$[-1,1]$. Suppose we
modify the chain by including self-loops at every state, i.e.,
remaining at the current state with probability~$1/2$, and following
the transition of the chain with probability~$1/2$. Then the
transition matrix becomes~$(\mathbb{I} + P)/2$, where~$\mathbb{I}$ is
the identity matrix. The eigenvalues of this matrix lie in the
interval~$[0,1]$.  Because of such self-loops, all the eigenvalues of
our walk above on~$S_l$ are non-negative.

In order to find a lower bound for the spectral gap of our random walk
on~$S_l$, we use Corollary~\ref{thm-gap-coupling-time}.
A coupling for which~$T \le l \log l$ is the obvious one: for any
pair~$u,v \in S_l$, follow one step of the random walk with the same
choice of random position~$i$ and index~$j$. This is clearly a valid
coupling, since the marginal process on any one of the two tuples is the
same as our walk, and if the two tuples are identical, they are modified
identically by the walk.

Let~$d$ be the Hamming distance between the two tuples~$u,v$.  This
distance never increases during the coupling process described above.
Moreover, in one step of the process, the distance goes down by~$1$ with
probability at least~$\frac{d}{2l}$. This is because with
probability~$d/l$, the position~$i$ is one where~$u$ and~$v$ are
different, and with probability at least~$(k-l)/k$, the index~$j$ is
not one from the positions where~$u$ and~$v$ are the same. Since~$l
\le k/2$, the net probability that the distance decreases by~$1$ is at
least~$d/2l$.

By a straightforward calculation, the expected time~$T$ for the
distance to go to zero is at most~$2l \log l$ (since~$d \le l$).
Using the relation between~$\lambda$ and~$T$ derived in
Corollary~\ref{thm-gap-coupling-time}, we get
our bound on the spectral gap.
\end{proof}

\begin{theorem}\label{algo-thm}
There is a quantum algorithm that solves \textsc{Group Commutativity}
with~$O( k^{2/3} \log k)$ queries 
and time
complexity $O( k^{2/3} \log^2 k)$.
\end{theorem}
\begin{proof}
Our algorithm derived from an application of 
Theorem~\ref{mario-thm} to the product of two independent walks
on~$S_l$.  The database associated with a tuple~$u\in S_l$ is the
binary tree $t_u$. Due to the Szegedy Theorem, we need only compute
the eigenvalue gap~$\delta$ of the random walk and the fraction of
marked states~$\varepsilon$ in the uniform distribution on~$S_l^2$.

The stationary distribution for the walk is the uniform distribution
on~$S_l \times S_l$.  So, from Lemma~\ref{com-lemma} above, the
fraction of marked states~$\varepsilon$ is at least $(1-p)^2/4$. The
spectral gap~$\delta$ for the walk is the same as that on~$S_l$,
i.e.,~$\delta \ge c/(l \log l)$, from Lemma~\ref{thm-gap}.

We start with a uniform distribution over~$\ket{u,t_u}\ket{v,t_v}$,
where $u,v\in S_{l}$.  The setup cost is at most $2(l-1)$ and the
updating cost of the walk is~$O(\log l)$. 
We choose $l=o(k)$ so
that $1-p=\Theta(l/k)$.  By Theorem~\ref{mario-thm}, the total query cost is
\begin{eqnarray*}
\lefteqn{ 2(l-1) + O\left( \frac{1}{\sqrt{\delta\varepsilon}} \log l \right) } \\
  & = & 2(l-1) + O \left( \frac{1}{1-p} \sqrt{l \log l} \cdot \log l \right) \\
  & = & 2(l-1) + O \left( \frac{k}{\sqrt{l}} \log^{3/2}l \right).
\end{eqnarray*}
This expression is minimized when~$l = k^{2/3} \log k$, and the cost
is~$O(k^{2/3} \log k)$.

The time complexity overhead comes from the initialization and
transition times that are both essentially equal to the time
complexity of performing a Grover diffusion operation (see
Section~\ref{sec-qwalks}).  For the initialization, we use a diffusion
over $S_l^2$, whose time complexity is $O(\log (\size{S_l}^2))=O(l\log
k)$.  For the transition, we use a diffusion over a set of size $2$
tensor product with a diffusion over a set of size $kl$, therefore the
corresponding time complexity is $O(\log (kl))=O(\log k)$.
\end{proof}

\section{Reduction from \usc}
\label{sec-lb}

We begin our presentation of the lower bound by considering the
complexity of~\usc.  This problem is at least as hard as \uc\ in its
query complexity since any bounded-error algorithm for the former can
be used to detect an arbitrary collision. 

\begin{proposition}\label{usc-prop}
The randomized and the quantum query complexity  of \usc\ is respectively
$\Omega(k)$ and $\Omega(k^{2/3})$. 
\end{proposition}
\journal{\begin{proof} One can prove the $\Omega(k)$ lower bound for
classical query complexity by an adversary argument.

For the quantum case, we reduce~\uc\ to~\usc\ by composing the oracle
function with a random permutation.  Then Theorem~\ref{ed-thm}
(due to Aaronson and Shi~\cite{as04} and Kutin~\cite{Kutin05}, together with
Ambainis~\cite{Ambainis05})
implies  the lower bound.

Assume that we have an algorithm $A$ for~\usc\ with constant bounded
error $\gamma<1/4$. We run~$A$ on oracle~$F$ composed with a random
permutation on the domain.  If there is a collision in the function
$F$, with probability at least~$1/2$, the colliding pair will have one
point on either side of~$k/2$. This will be detected with probability
at least~$1-\gamma$. The overall success probability will
be~$(1-\gamma)/2 \ge 3/8$. If there is no collision, the algorithm
will make an error with probability at most~$\gamma < 1/4$.  Using
standard techniques, this gap in acceptance probability can be made
symmetric around~$1/2$ and boosted by repeating the experiment with an
independent run of the algorithm~$A$ on~$F$ composed with a fresh
random permutation. For completeness, we include the argument below.

Our final algorithm for \uc\ picks two random permutations, and
runs $A$ on the oracle function composed with these permutations. It
accepts if any one of the two executions of $A$ accepts.  We now show
that the error of our algorithm is now upper bounded by
$1/4+\gamma<1/2$.

If the oracle function $F$ has no collision, the error is upper
bounded by $2\gamma < 1/4+\gamma$.  If $F$ has a collision, then with
probability at most $(1/2)^2=1/4$, the colliding pair has no point on
either side of $k/2$ for both the randomly chosen permutations.
Assume this is not the case, and fix a permutation for which the
permuted $F$ has a split collision.  Our algorithm accepts this
permutation of $F$ with probability at least $1-\gamma$.  Therefore
the overall error is upper bounded by $1/4+\gamma$.

In reducing a positive instance of~\uc, with probability close
to~$1/4$, we get inputs for which the algorithm for~\usc\ need not
output a definite answer with probability bounded away
from~$1/2$. (These are inputs where the colliding pair of indices are
both at most~$k/2$ or both greater than~$k/2$.) Our argument is valid
in spite of this, since the acceptance probability in the other case
is high.
\end{proof}}

We conclude by proving the same lower bound for \textsc{Group
Commutativity} as well.  We thus show that the algorithm described in
the previous section is almost optimal.

The group involved in the proof of the lower bound is a subgroup
$G$ of~$U(4k)$, the group under matrix multiplication of~$4k \times
4k$ unitary matrices. The generators of $G$ are block diagonal,
each with~$2k$ blocks of dimension~$2 \times 2$. Each block is one
of the following Pauli matrices:
\[
I  =  \left(\begin{array}{cc} 1 & 0 \\ 0 & 1 \end{array} \right),\quad
X  =  \left(\begin{array}{cc} 0 & 1 \\ 1 & 0 \end{array} \right), \quad
Z  =  \left(\begin{array}{cc} 1 & 0 \\ 0 & -1 \end{array} \right), \quad
Y = XZ = \left(\begin{array}{cc} 0 & -1 \\ 1 & 0 \end{array} \right).
\]
No pair of matrices amongst~$X,Y$ and~$Z$ commute. An encoding of the
group~$G$ consists in words $\sigma_1\ldots \sigma_{2k}$ of length
$2k$ over the alphabet $\{I,X,Y,Z\}$ together with a sign vector~$s =
(s_1,s_2,\ldots,s_{2k})$ in $\{+1,-1\}^{2k}$. A tuple
$(s,\sigma_1,\ldots \sigma_{2k})$ represents the matrix
$\mathrm{diag}(s_1 \sigma_1,\ldots, s_{2k} \sigma_{2k})$.  We 
call this encoding the {\em explicit encoding}.

\begin{theorem}\label{slb-thm}
The randomized and the quantum query complexity of \textsc{Group
Commutativity} are respectively $\Omega(k)$ and $\Omega(k^{2/3})$.
\end{theorem}
\begin{proof}
We prove the theorem by reducing \usc\ to \textsc{Group
Commutativity}. First, we define a group that is non-commutative if and only
if the oracle input~$F$ for \usc\ has a collision. Second, we design
a unique encoding of the group elements such that each group operation can be
simulated with at most four queries to~$F$.
We then conclude our theorem using Proposition~\ref{usc-prop}.

For~$i,j \in \set{1,2, \ldots, k}$, we define $a_{ij}, b_{ij} \in
U(2k)$ of the form described above. Both kinds of matrix have the
identity matrix in all their blocks except for the~$i$-th and
$(j+k)$-th.  The $i$-th block is $Y$ in both~$a_{ij}$ and~$b_{ij}$.
The~$(j+k)$-th block is~$Z$ in~$a_{ij}$ and~$X$ in~$b_{ij}$.

Suppose the oracle for the problem~\usc\ computes the
function~$F : \set{1,\ldots,k} \rightarrow \set{1,\ldots,k}$. We
associate a generator~$g_i$ of the type described above with each
element~$i$ in the domain of~$F$. The generator~$g_i$ is $a_{iF(i)}$
if~$i \le k/2$, and it is~$b_{iF(i)}$ if~$i > k/2$.

Observe that all the generators~$g_i$ are distinct, since the~$i$-th
block in it is~$Y$, and the rest of the initial~$k$ blocks are the
identity matrix. This is designed so that we can identify the
index~$i$ from the explicit encoding of a generator.

For any two distinct points~$i_1,i_2$, if~$F(i_1) \not= F(i_2)$, then
the generators~$g_{i_1}, g_{i_2}$ have distinct blocks~$F(i_1)+k$
and~$F(i_2)+k$ set to either~$Z$ or~$X$. So if the function~$F$ is
injective, the set of generators~$\set{g_i}$ consists of~$k$ distinct
commuting elements.  If there is a collision~$i_1,i_2$ in~$F$ with one
point on either side of~$k/2$, then the same block~$F(i_1)+k =
F(i_2)+k$ is set to~$Z$ and~$X$ respectively. 
Then the generators~$g_{i_1}$ and~$g_{i_2}$ do not commute, and
the group generated by~$\set{g_i}$ is non-abelian.


The encoding of the group elements is the explicit encoding 
defined above except for the generators $g_i$. The generators~$g_i$ 
are encoded by their corresponding indices $i$. The input to \gc\  
is~$1,2,\ldots, k$.

Now we explain how to simulate the group operations.
When an integer $i$ is involved in a group operation, we query the oracle
for~$F$ at~$i$ and construct~$g_i$ as defined above. One more query
to~$F$ is required to erase the value of the function.
Otherwise we do not query~$F$. Matrix
operations can be performed without incurring any further calls
to~$F$.

We also have to take care to output the result of a group operation in
the correct encoding.  Namely, when the output of a group operation is
either $a_{ij}$ or $b_{ij}$, for some~$i,j$, then we check if it can
be encoded by~$i$ using one query to~$F$, and one more query to erase
the value of the function.

Thus a group operation involves~$F$ at most six times, when both of
the elements are encoded by integers.  Note that a product of two
group elements of the form~$a_{ij}$ or~$b_{ij}$ can never result in
another such generator. We can thus improve the number of invocations
of the function oracle from six to four per group operation.
\end{proof}

\section{Acknowledgements}

We thank the anonymous referees for ICALP'05 and {\it Algorithmica\/}
for their helpful feedback on the paper.


\begin{thebibliography}{BBC{\etalchar{+}}01}

\bibitem[Ald82]{Aldous82}
David Aldous.
\newblock Random walks on finite groups and rapidly mixing {M}arkov chains.
\newblock In {\em S{\'e}minaire de Probabilit{\'e}s XVII}, volume 986 of {\em
  Lecture Notes in Mathematics}, pages 243--297. Springer-Verlag, 1981--82.

\bibitem[Amb04]{amb04}
Andris Ambainis.
\newblock Quantum walk algorithm for {Element Distinctness}.
\newblock In {\em Proceedings of the 45th Annual IEEE Symposium on Foundations
  of Computer Science}, pages 22--31. IEEE Computer Society Press, Los
  Alamitos, CA, 2004.

\bibitem[Amb05]{Ambainis05}
Andris Ambainis.
\newblock Polynomial degree and lower bounds in quantum complexity: {Collision}
  and {Element Distinctness} with small range.
\newblock {\em Theory of Computing}, 1(3):37--46, 2005.

\bibitem[AS04]{as04}
Scott Aaronson and Yaoyun Shi.
\newblock Quantum lower bounds for the collision and the element distinctness
  problems.
\newblock {\em Journal of the ACM}, 51(4):595--605, 2004.

\bibitem[BBC{\etalchar{+}}01]{bbcmw01}
Robert Beals, Harry Buhrman, Richard Cleve, Michele Mosca, and Ronald de~Wolf.
\newblock Quantum lower bounds by polynomials.
\newblock {\em Journal of the ACM}, 48(4):778--797, 2001.

\bibitem[BS84]{bs84}
L{\'a}szl{\'o} Babai and Endre Szemer\'edi.
\newblock On the complexity of matrix group problems {I}.
\newblock In {\em Proceedings of the 25th Annual IEEE Symposium on Foundations
  of Computer Science}, pages 229--240, 1984.

\bibitem[Bv06]{BuhrmanS06}
Harry Buhrman and Robert \v{S}palek.
\newblock Quantum verification of matrix products.
\newblock In {\em Proceedings of the Seventeenth Annual ACM-SIAM Symposium on
  Discrete Algorithms}, pages 880--889, 2006.

\bibitem[Gri78]{Griffeath78}
David~S. Griffeath.
\newblock Coupling methods for {M}arkov processes.
\newblock In Gian-Carlo Rota, editor, {\em Studies in Probability and Ergodic
  Theory}, pages 1--43. Academic Press, New York, NY, USA, 1978.

\bibitem[Gro96]{gro96}
Lov~K. Grover.
\newblock A fast quantum mechanical algorithm for database search.
\newblock In {\em Proceedings of the Twenty-Eighth Annual {ACM} Symposium on
  the Theory of Computing}, pages 212--219, 1996.

\bibitem[KSV02]{ksv02}
Alexei~Yu. Kitaev, Alexander~H. Shen, and Mikhail~N. Vyalyi.
\newblock {\em Classical and Quantum Computation}, volume~47 of {\em Graduate
  Studies in Mathematics}.
\newblock American Mathematical Society, Providence, RI, 2002.

\bibitem[Kut05]{Kutin05}
Samuel Kutin.
\newblock Quantum lower bound for the collision problem with small range.
\newblock {\em Theory of Computing}, 1(2):29--36, 2005.

\bibitem[Mos99]{mos99}
Michele Mosca.
\newblock {\em Quantum Computer Algorithms}.
\newblock PhD thesis, University of Oxford, 1999.

\bibitem[MSS05]{mss05}
Fr{\'e}d{\'e}ric Magniez, Miklos Santha, and Mario Szegedy.
\newblock Quantum algorithms for the triangle problem.
\newblock In {\em Proceedings of 16th Symposium on Discrete Algorithms}, pages
  1109--1117. ACM-SIAM, 2005.

\bibitem[NC00]{nc00}
Michael~A. Nielsen and Isaac~L. Chuang.
\newblock {\em Quantum Computation and Quantum Information}.
\newblock Cambridge University Press, Cambridge, UK, 2000.

\bibitem[Pak00]{pak00}
Igor Pak.
\newblock Testing commutativity of a group and the power of randomization.
\newblock Electronic version at {\tt
  http://www-math.mit.edu/$\sim$pak/research.html}, 2000.

\bibitem[Sin93]{sinclair93}
Alistair Sinclair.
\newblock {\em Algorithms for Random Generation and Counting: A {M}arkov Chain
  Approach}.
\newblock Progress in theoretical computer science. Birkh\"{a}user, Boston,
  1993.

\bibitem[Sze04]{sze04}
Mario Szegedy.
\newblock Quantum speed-up of {M}arkov chain based algorithms.
\newblock In {\em {Proceedings of the 45th Annual IEEE Symposium on Foundations
  of Computer Science}}, pages 32--41, 2004.

\bibitem[Wat01]{wat01}
John Watrous.
\newblock Quantum algorithms for solvable groups.
\newblock In {\em Proceedings of 33rd Symposium on Theory of Computing}, pages
  60--67. ACM, 2001.

\end{thebibliography}

\newcommand{\etalchar}[1]{$^{#1}$}

\end{document}